\documentclass[12pt]{elsart}

\usepackage{epsfig,amsfonts,newlfont,rotate}      
\usepackage{amsmath,amssymb} 

\def\eq#1{(\ref{#1})}
\newcommand{\pardis}{\langle \mu \rangle} 
\newcommand{\Tc}{T_\mathrm{c}}
\newcommand{\bc}{\beta_\mathrm{c}}
\newcommand{\vi}{\vec{\mbox{{\it \i}}}}
\newcommand{\vj}{\vec{\mbox{{\it \j}}}}

\begin{document}
\runauthor{Carmona, Di Giacomo and Lucini}
\begin{frontmatter} 
\title{A disorder analysis of the Ising model}
\author[Pisa]{J.M. Carmona\thanksref{mail1}},
\author[Pisa]{A. Di Giacomo\thanksref{mail2}},
\author[Oxford]{B. Lucini\thanksref{mail3}}
\address[Pisa]{Dipartimento di Fisica dell'Universit\`a
and INFN, I-56127 Pisa (Italy)}
\address[Oxford]{Theoretical Physics, University of Oxford,
1 Keble Road, Oxford, OX1 3NP (UK)}
\thanks[mail1]{E-mail: carmona@df.unipi.it}
\thanks[mail2]{E-mail: digiaco@df.unipi.it}
\thanks[mail3]{E-mail: lucini@thphys.ox.ac.uk}

\begin{abstract}
Lattice studies of monopole condensation in QCD are based on the
construction of a disorder parameter, a creation operator of monopoles
which is written in terms of the gauge fields. This procedure is
expected to work for any system which presents duality. We check it
on the Ising model in $2d$, which is exactly solvable. The output
is an amusing exercise in statistical mechanics.
\end{abstract}

\begin{keyword}
Lattice; Monte Carlo; Order-disorder;
Finite size scaling; Ising model
\end{keyword}

\end{frontmatter}

\section{Introduction}
Duality~\cite{kramers,kadanoff} is a property of many systems in 
field theory and in \mbox{statistical}
mechanics. Their common feature is that, when viewed as Euclidean
field theories, they have spatial [$(d-1)$-dimensional] configurations
carrying a conserved topological charge. The high temperature (strong
coupling) phase is characterized by condensation of these topological
structures. In a dual description these structures are described by
local fields, and their v.e.v. acts as a dual order parameter, or
disorder parameter.

In a few models the transformation to the dual is known analytically
(the Ising model, the U(1) model with Villain action~\cite{marchetti},
the $3d$ XY model). In other theories duality is expected to be at
work, but its presence has only been tested
numerically~\cite{deldebbio1,deldebbio2,paffuti,mario,dario,digiaco}.
The idea is to guess the topological symmetries classifying 
the extended structures, construct a disorder
parameter as the creation operator of a structure with non-trivial 
topology~\cite{swieca}, and then measure it by numerical 
simulations~\cite{deldebbio1,deldebbio2,paffuti,mario,dario,digiaco}.

The approach~\cite{swieca,deldebbio1,deldebbio2,paffuti,mario,dario,digiaco}
relies on symmetry, and is insensitive to the
exact choice of the action, or to irrelevant terms, contrary to the
explicit transformation to the dual, 
which is only possible for specific
forms of the action (the Villain action in the XY model 
and in the U(1) model in $(3+1)d$).

A test of the construction is the determination of the critical indices
at the phase transition by finite size scaling analysis. This 
has already been done for QCD~\cite{digiaco}, where the natural 
topological structures in $3d$ are monopoles. A clear evidence has
emerged of dual superconductivity as the mechanism of colour
confinement.

In $(2+1)$-dimensional models the topological structures are vortices
(XY model) or $2d$ O(3) instantons (Heisenberg model). The XY model in
its Villain form was already known to present duality and condensation 
of vortices in the strong coupling phase. Numerical studies by
the symmetry approach confirmed that. The Heisenberg model was
first viewed as a dual system in Ref.~\cite{dario}. 

In this paper we test the symmetry approach on the simplest and 
prototype model in $(1+1)d$ presenting duality: the Ising model.
We show as a theorem that our construction of the disorder parameter
is identical to that of Ref.~\cite{kadanoff}. We give an explicit
construction of the dual variables, and we check our numerical
approach vs the exact solution of the model. 

\section{Disorder parameter}
\label{sez2}
The action of the $2d$ Ising model is
\begin{equation}
S[\sigma]=-J \sum_{\vec n,\mu} \sigma(\vec n)
\sigma(\vec n +\hat\mu) \ ,
\label{isingaction}
\end{equation}
where $\vec n=(n_0,n_1)$ is the coordinate of the lattice site and 
$\hat\mu$ is the unit vector in the direction $\mu=0,1$.
$\sigma(\vec n)$ is the field variable, with values $\sigma(\vec n)=\pm 1$.
 
The partition function is
\begin{equation}
Z = \sum_{\{ {\sigma}(\vec n)\}} \exp \left[-\frac{1}{T} S[\sigma] \right] \ .
\label{zetaising0}
\end{equation}

The model is exactly solvable~\cite{onsager} 
and presents a second order
phase transition at the critical temperature $\Tc$
\begin{equation}
\Tc = \frac{2J}{\ln (\sqrt{2}+1)} \ .
\end{equation}

The order parameter is the magnetization
\begin{equation}
\langle\sigma\rangle = \lim_{V\to\infty} \frac{1}{V}
\sum_{\vec n} \langle\sigma(\vec n)\rangle \ .
\end{equation}
At low temperatures $T<\Tc$, $\langle\sigma\rangle\neq 0$ and the symmetry 
of the action under the transformation
\begin{equation}
\sigma \to -\sigma
\end{equation}
is spontaneously broken. For $T>\Tc$, $\langle\sigma\rangle=0$ (in the
infinite volume limit $V\to\infty$).

The critical exponent known as $\beta$, which governs the behaviour of
$\langle\sigma\rangle$ near $\Tc$,
\begin{equation}
\langle\sigma\rangle \propto \left( \Tc - T \right)^{\beta} \ ,
\end{equation}
has the value
\begin{equation}
\beta = \frac{1}{8} \ .
\label{betaexp}
\end{equation}  

The correlation function of the field $\sigma$
\begin{equation}
\Gamma(\vi,\vj\,) = \langle \sigma(\vi\,) \sigma(\vj\,) \rangle 
\end{equation}
near the phase transition is approximately rotation invariant, only
depends on the distance $d=\sqrt{(\vi-\vj\,)^2}$ and behaves as
\begin{equation}
\Gamma \sim \exp\left(-\frac{d}{\xi(T)}\right) + \langle\sigma\rangle^2 \ .
\end{equation}
For $T>\Tc$, $\langle\sigma\rangle=0$, and the correlation length $\xi(T)$
diverges at $T=\Tc$ as
\begin{equation}
\xi = a \left(T - \Tc \right)^{- \nu} \ .
\end{equation}
The critical index $\nu$ has the value $\nu=1$.

A dual description of the system can be given in terms of a dual field
$\mu$, defined on the dual lattice~\cite{kadanoff}. The dual lattice
associates to each plaquette of the original lattice a site, which can 
be identified as its centre, and to each link a dual link, which is 
orthogonal to it and cuts it. The variable $\mu$ is defined throughout
its correlation functions
\begin{equation}
\langle \mu(\vi\,)\mu(\vj\,)\rangle = \frac{Z[\bar S]}{Z[S]} \ .
\end{equation}
$\bar S$ differs from $S$ by the introduction of a magnetic
dislocation, i.e. by changing signs to $J$ on all the links crossed
by a line joining $\vi$ and $\vj$ on the dual lattice: the
result is independent of the choice of the line.

The field $\mu$ obtained in this way is again valued $\mu=\pm 1$ and
the partition function $Z[\sigma,T]$ obeys the equality
\begin{equation}
Z[\sigma,T]=Z[\mu,T^*] \ ,
\end{equation}
with 
\begin{equation}
\sinh\frac{2}{T}=\frac{1}{\sinh\frac{2}{T^*}} \ ,
\end{equation}
i.e. $T^*\gtrless\Tc$ if $T\lessgtr\Tc$.

The dual description of the system maps the disordered phase
$\langle\sigma\rangle=0$ onto the ordered phase of the dual.
For $T>\Tc$, $\langle\mu\rangle\neq 0$, $\langle\sigma\rangle=0$. The
order parameter of the disordered phase is called a 
{\em disorder parameter.} Near the critical point
\begin{equation}
\langle\mu\rangle \propto \left( T - \Tc \right)^{\delta} \ ,
\end{equation}
which defines the critical exponent $\delta$ associated to the
disorder parameter. Then, self-duality of the model implies that
\begin{equation}
\delta=\beta=\frac{1}{8} \ .
\end{equation}

We shall approach the problem in a slightly different form, 
emphasizing the symmetry aspects of the disordered phase. We shall then
prove that our disorder parameter is equal to that of Ref.~\cite{kadanoff}.
A similar procedure was used in Ref.~\cite{paffuti} for the U(1)
gauge theory, as an alternative procedure to that of Ref.~\cite{marchetti}.

The Ising model can be viewed as a $1+1$ dimensional Euclidean field
theory, with action
\begin{equation}
S= -\frac{J}{2} \sum_{\mu=0,1}\sum_{\vec n}
[\Delta_\mu \sigma(\vec n)]^2 \ ,
\end{equation}
with $\Delta_\mu\sigma(\vec n)=\sigma(\vec n+\hat\mu)-\sigma(\vec n)$.
The current
\begin{equation}
j_\mu=\frac{1}{2}\epsilon_{\mu\nu}\Delta_\nu \sigma
\end{equation}
is identically conserved,
\begin{equation}
\Delta_\mu j_\mu=0 \ .
\label{conservlaw}
\end{equation}
The corresponding constant of motion is
\begin{equation}
Q=\sum_{n_1} j_0(n_0,n_1)
\label{charge}
\end{equation}
or
\begin{equation}
Q=\frac{1}{2}\sum_{n_1} \Delta_1 \sigma(n_0,n_1) =
\frac{1}{2} \left[\sigma(n_0,+\infty)-\sigma(n_0,-\infty)\right] \ .
\end{equation}

$Q$ is a topological charge, which classifies 1-dimensional spatial
configurations by their boundary conditions. A {\em kink} corresponds
to $\sigma(n_0,+\infty)=1$, $\sigma(n_0,-\infty)=-1$ or $Q=1$, an
antikink to $Q=-1$.

Our guess is that the conservation law Eq.~\eq{conservlaw} is
spontaneously broken in the high-temperature phase, by condensation
of kinks. As a disorder parameter we shall choose the vacuum
expectation value of an operator carrying nonzero charge $Q$ 
(Eq.~\eq{charge}). For that operator we choose the creation operator
of a kink.

The creation operator of a kink at site $n_1$ and time $n_0$ is
defined as 
\begin{equation}
\mu(n_0,n_1)=\exp\left[-2\,\frac{J}{T}\sum_{n\leq n_1} \sigma(n_0,n)
\sigma(n_0+1,n)\right] \ .
\end{equation}
The correlator
\begin{equation}
{\mathcal D}(n_0) = \langle\mu(n_0,n_1)\mu(0,n_1)\rangle
\label{mucorrelator}
\end{equation}
is accordingly defined as
\begin{equation}
{\mathcal D}(n_0)= \frac{1}{Z}\sum_{\{\sigma\}}
\exp\left(-\frac{S}{T}\right) \mu(n_0,n_1)\mu(0,n_1)
=\frac{\bar Z}{Z} \ .
\end{equation}
$\bar Z$ is obtained from $Z$ by reversing the sign of the
temporal links $J\sigma(0,n)\sigma(1,n)$ and 
$J\sigma(n_0,n)\sigma(n_0+1,n)$, $n\leq n_1$,  in the action.

It is trivial to check that ${\mathcal D}(n_0)$ Eq.~\eq{mucorrelator}
really corresponds to the propagator of a kink at the spatial site
$n_1$ from time $t=0$ to time $t=n_0$. Indeed in computing
$\bar Z$ we can change variables in the sum from
$\sigma(1,n)$ to $-\sigma(1,n)$, $n\leq n_1$. This change brings
the temporal links 0-1 to 
the original form they appeared in $Z$, but
changes sign to the spatial link $\sigma(1,n_1)\sigma(1,n_1+1)$
and to the temporal links $\sigma(1,n)\sigma(2,n)$, $n\leq n_1$. A
kink has been added to the configuration at time $n_0=1$. We can
now change variables by sending $\sigma(2,n)\to -\sigma(2,n)$. The
result will be again the appearance of a kink at $n_0=2$, and 
a change of sign of links $\sigma(2,n)\sigma(3,n)$, $n\leq n_1$.
The construction can be repeated and finally the change of sign
of the temporal link at $n_0$ will be reabsorbed by the kink
sitting at $n_0$ in $\bar Z$. This completes the proof.

The net effect is a change of the spatial links 
$(n_1,n_1+1)$ at all times in the interval
$(0,n_0]$. This is exactly the definition of the dual 
correlator given in Ref.~\cite{kadanoff}. So our $\langle\mu\rangle$ is 
equal to the disorder parameter of
Ref.~\cite{kadanoff} and provides an explicit
construction of the operator in terms of the original 
fields $\sigma$. The construction is exactly the same which was
given in Refs.~\cite{paffuti,mario,dario}, respectively 
for the U(1), the $3d$ XY model and the Heisenberg model.

Some remarks are necessary at this point.
First, in the absence of the second kink at $n_0$, the construction sketched 
above includes
a change of sign of the temporal link at the border of the lattice
and the periodic boundary conditions are made
antiperiodic by the change of variables. 
Second, for finite lattices
the correlator Eq.~\eq{mucorrelator} is a four point function.
As $n_0$ goes large, by cluster property
\begin{equation}
{\mathcal D}(n_0) \simeq \langle\mu\rangle^4 \ .
\end{equation}
Finally, we stress that the result of our construction
is mathematically identical to that of Ref.~\cite{kadanoff}.
$\langle\mu\rangle\neq 0$, $T>\Tc$ is a theorem. However for more
complicated choices of the action, belonging to the same
universality class, our construction, which only relies on symmetry,
can be more practical for numerical approaches. This is 
similar to the situation for U(1), where the construction of the 
dual is feasible for the Villain form of the action, but the 
approach based on symmetry allows treatment of Wilson and other 
forms of the action.

As a check of our arguments we shall compute the disorder parameter
numerically, and use it to determine the critical indices of the
model.

\section{Numerical simulations}
\label{sez3}
Let us try to extract information about the phase transition from the
numerical study of the disorder parameter. To this purpose, at fixed $n_1$ 
in a finite lattice with periodic boundary conditions, we study 
\begin{equation}
\label{2ising}
{\cal D}(n_0) = \frac{\bar{Z}}{Z} \ 
\end{equation}
as a function of $\beta= 1/T$ by standard Monte Carlo methods.

By the cluster property, at large time separation it is expected that
\begin{equation}
{\cal D}(n_0) \to \pardis ^4 \ ,
\label{clusterprop}
\end{equation}
being $\pardis$ the disorder parameter. Hence in principle we can
extract $\pardis$ by a fit of ${\cal D}(n_0)$ data
at large $n_0$. $\pardis$ is expected to have the opposite behaviour
to the magnetization $M$: in the thermodynamic
limit $\pardis$ is rigorously zero for $\beta > \bc$, 
different from zero for $\beta < \bc$, and for $\beta \lesssim \bc$,
it approaches zero with the power law
\begin{equation}
\label{disscaling}
\pardis \simeq t ^{\delta} \ ,
\end{equation}
where $t$ is the reduced temperature $(T - \Tc)/\Tc$.

A direct determination of ${\cal D}(n_0)$ by Monte Carlo techniques
requires an unbearable computational cost.
This is a consequence of the form of ${\cal D}(n_0)$ in terms of the order
variables \cite{deldebbio1}. 
In fact ${\cal D}(n_0)$ is the ratio of two partition
functions and the determination of the partition function of a system by
numerical simulations is very difficult: $Z$ is the exponential of a quantity
fluctuating like the square root of the volume, so we expect for it
fluctuations of order $e^{\sqrt{V}}$. Because of these strong fluctuations
the direct determination of ${\cal D}(n_0)$ by Monte Carlo techniques
requires an amount of statistics which can not be obtained in realistic 
simulations.

This difficulty can be overcome by defining another quantity whose
fluctuations are proportional to the square root of
the volume and which gives information
about the behaviour of $\pardis$.
To this goal we introduce the function \cite{deldebbio1}
\begin{equation}
\rho = \frac {\d }{\d \beta} \log {\cal D}(n_0) \ .
\end{equation}
In the limit $n_0 \to \infty$
\begin{equation}
\rho \to 4 \frac {\d} {\d \beta} \log \pardis \ ,
\end{equation}
which implies
\begin{equation}
\pardis = \exp \left( \frac{1}{4}
\int _0 ^{\beta} \rho(\beta^{\prime}) \d
\beta^{\prime} \right) \ ,
\end{equation}
and the behaviour of $\pardis$ can be reconstructed from the behaviour
of $\rho$ in a simple way.

By using Eq.~(\ref{2ising}) we get
\begin{equation}
\label{rhoisingex}
\rho = \langle S \rangle _S - \langle \bar{S} \rangle _{\bar{S}} \ ,
\end{equation}
where $\langle S \rangle _S$ is the average of the action of the Ising model,
Eq.~(\ref{isingaction}), taken on configurations generated with that action,
and $\langle \bar{S} \rangle _{\bar{S}}$ is the average of the
{\em kink action\/} $\bar{S}$ taken on configurations generated with this
modified action. The quantity $\rho$ will be much easier to measure in
a numerical simulation, since it is expected to fluctuate as $\sqrt{V}$.

\begin{figure}[tb]
\begin{center}
\epsfig{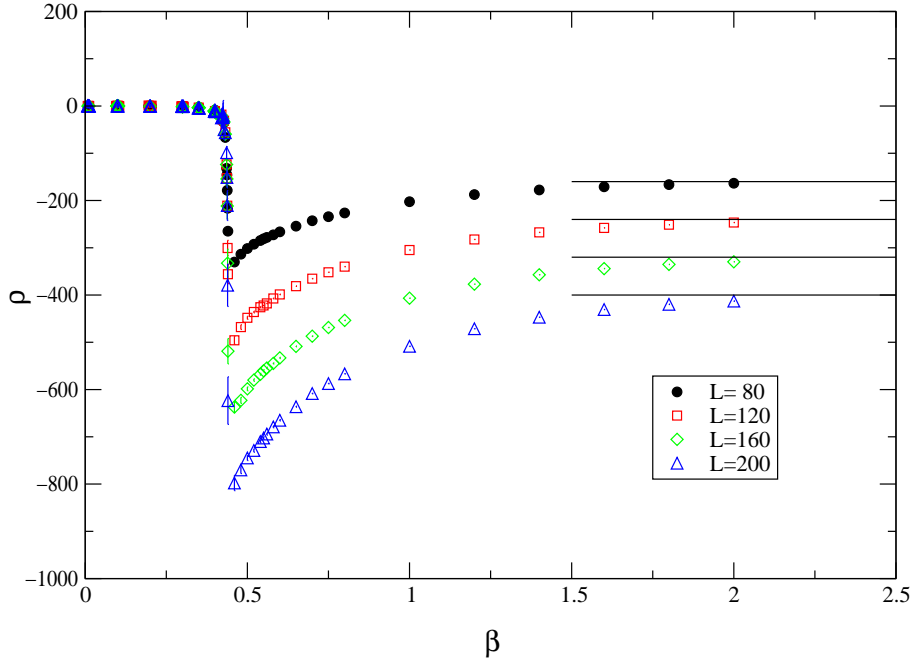} 
\end{center}
\caption{$\rho$ as a function of $\beta$ for different lattice sizes. 
Continuous lines refer to low temperature calculations.} 
\label{rhoising.fig}
\end{figure}

We have investigated lattices of size $L\times 2L$, with the spatial 
extension $L$ ranging from 80 to 200. In our
simulations we chose $n_1 = L/2$ and $n_0 = L$, and periodic boundary
conditions were imposed. We simulated both the standard and the kink 
action of the model by means of a heat-bath algorithm. Errors
were estimated by applying the jack-knife method to bunched data.
Far from criticality we collected about
100000 measurements for each $\beta$ by sampling each two sweeps; near
the critical coupling the statistics has been increased by a factor of
25.  

From the behaviour of $\pardis$ in the thermodynamic limit,
we expect that $\rho$ is nearly constant at low $\beta$ (due to the slow
variation of $\pardis$ in this region), has a sharp negative peak in the
critical region (corresponding to the sudden decline of $\pardis$) and
approaches $- \infty$ at weak coupling (as a consequence of the restoring of
the dual symmetry). The extrapolation of numerical results to the 
thermodynamic limit should give this behaviour for $\rho$.
A plot of $\rho$ data for the lattices we have investigated is reported in
Fig.~\ref{rhoising.fig}. We shall show that the obtained
shape of $\rho$ is compatible
with the expected behaviour in the infinite volume limit.  

By simple algebraic manipulation,
\begin{equation}
{\cal D}(n_0) = \langle \e^{ - \beta (\bar{S} - S)} \rangle _S \ ,
\end{equation}
where the subscript $S$ means that the average has to be taken over
configurations weighted with the standard Ising Boltzmann factor.
Only modified links (i.e. links whose sign has been changed)
contribute to $\bar{S} - S$, so we get
\begin{equation}
\label{dstrong}
{\cal D}(n_0) = \langle \e^{ - 2 \beta \sum _{l^{\prime}} \sigma_i \sigma_k}
\rangle _S \ ,
\end{equation}
where $\sigma_i$ and $\sigma_k$ are spins connected by a modified
link $l^{\prime}$ and the sum is performed over all modified links.

\begin{figure}[tb]
\begin{center}
\epsfig{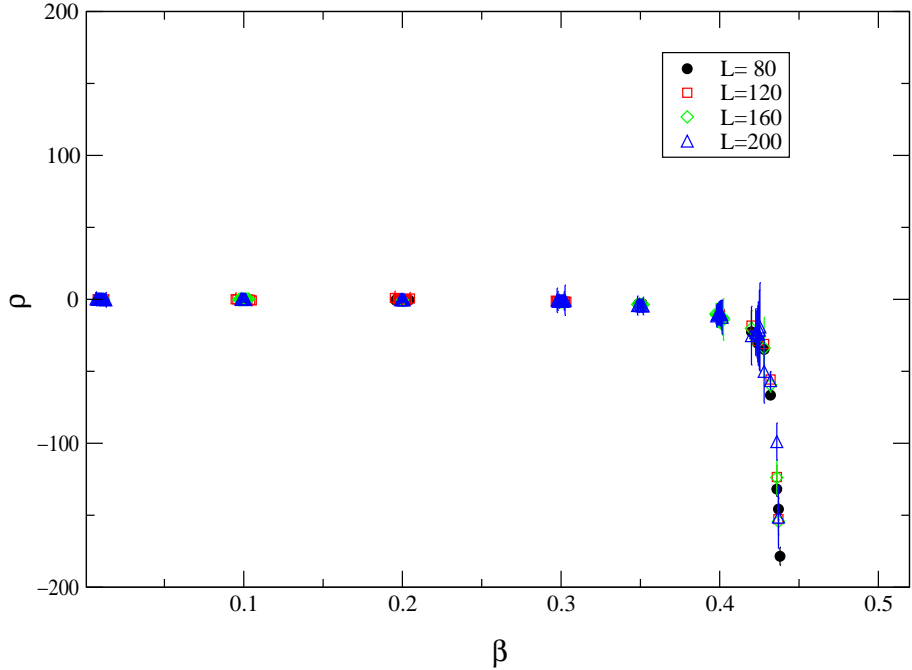} 
\end{center}
\caption{Low $\beta$ data for $\rho$.}
\label{isingstrong.fig}
\end{figure}

In the limit $\beta \to \infty$ the system is completely ordered, so all
links are equal to 1 and the sum in Eq.~(\ref{dstrong}) simply reduces
to the total number of modified links, i.e.
\begin{equation}
\label{dstrong1}
{\cal D}(n_0) \to \e^{ - 2 \beta L} \ .
\end{equation}
Indeed at high $\beta$, $\pardis$ goes to zero in the infinite volume limit,
as required from the dual symmetry restoration.

From Eq.~({\ref{dstrong1}}) we get that
\begin{equation}
\label{rhoweakisi}
\rho \to - 2 L
\end{equation}
at high $\beta$. In Fig.~{\ref{rhoising.fig}} continuous lines indicate the
expected results from the high $\beta$ calculation. 
The agreement with numerical
results is evident. We can also consider the opposite implication: if
Eq.~(\ref{rhoweakisi}) holds in the weak coupling,
in the thermodynamic limit $\pardis = 0$ in this region.

On the other hand, $\rho$ is compatible with zero for a wide range of 
$\beta$'s below $\bc$, independently of the volume (see 
Fig.~\ref{isingstrong.fig}).
This implies that $\pardis$ is nearly
constant in this region and that feature seems to extrapolate to the
thermodynamic limit. Since at high temperature
\begin{equation}
\pardis = 1 \ ,
\end{equation}
our data for $\rho$ imply that $\pardis$ in the thermodynamic limit is
different from zero in a wide range of $\beta$'s below $\bc$. 
Hence the behaviour of $\rho$
we obtained is compatible with dual symmetry breaking for $\beta < \bc$.

\begin{figure}[tb]
\begin{center}
\epsfig{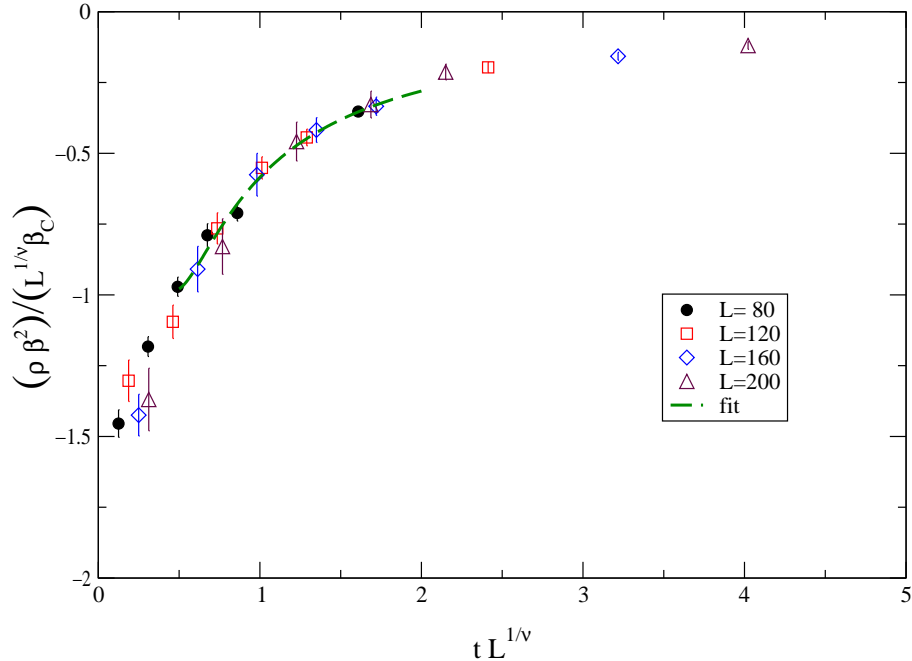} 
\end{center}
\caption{Rescaled plot of $\rho$ data.}
\label{scaising.fig}
\end{figure}

In the critical region the approach to the thermodynamic limit is governed
by finite size scaling theory, which states that at finite size $L$ in the
critical region for $T>\Tc$, $\pardis$ is given by
\begin{equation}
\label{scaising}
\pardis = t^{\delta} f( s) \ ,
\end{equation}
with $f$ an unknown function of the scaling variable
\begin{equation}
s = t L^{1/\nu} \ .
\end{equation}
From Eq.~(\ref{scaising}) it follows
\begin{equation}
\label{rhoisingsca}
\rho_{\textrm{rescal}}\equiv
\frac{\rho \beta^2}{L^{1/\nu} \bc} = - \frac{ 4 \delta}{s} - 
\frac{4}{f(s)} \frac{\d}{\d s} f(s) \ ,
\end{equation}
i.e. the rescaled variable $({\rho \beta^2})/({L^{1/\nu} \bc})$
is a function of the scaling variable $s$. 

A plot of rescaled
data is shown in Fig.~{\ref{scaising.fig}}. According to 
Eq.~\eq{rhoisingsca}, all points should fall on one single curve.
We observe however deviations at small and large $s$. It is important
to understand the range of validity of Eq.~\eq{rhoisingsca}.
On the one hand, in order for Eq.~\eq{clusterprop} to be valid, the 
antikink should be placed at a large temporal distance from the kink.
This means that 
\begin{equation}
n_0=L\gtrsim\xi ,
\label{cond1}
\end{equation}
where $\xi$ is the correlation length at
a certain temperature. On the other hand, we want to be in the scaling
region where finite size scaling is valid, so $L/\xi$ should not take very
large values. In order to have an approximate idea, we computed the
correlation length in a large $L\times L$ lattice, $L=400$, at the 
values of $\beta$ used to obtain Fig.~\ref{scaising.fig}. The results
are shown in Table~\ref{table:xi}.

\begin{table}[tb]
\begin{center}
\begin{tabular}{ccc}\hline
$\beta$ & $\xi$ & Lattices with scaling \\
\hline 
0.432 & 29.1(3) & 80 \\
0.436 & 52.7(2) & 80,120,160 \\
0.437 & 67.6(2) & 80,120,160,200\\
0.438 & 92.3(1) & 80,120,160,200\\
0.439 & 138.9(2) & 120,160,200 \\
0.44 & 238.1(4) &  - \\
\hline
\end{tabular}
\end{center}
\caption{Correlation length in the $2d$ Ising model measured
in a $400\times 400$ lattice at the $\beta$ values used in 
Fig.~\ref{scaising.fig}.
The third column indicates the lattice sizes which are expected to
show scaling behaviour (see text).}
\label{table:xi}
\end{table}
                
The condition Eq.~\eq{cond1} means that, for example for the $L=80$ 
lattice, only the points of $\beta\lesssim 0.438$ from the list shown in
Table~\ref{table:xi} should give a proper scaling of $\rho$. On the
other hand, $L$ should not be much more larger than $\xi$, because
otherwise we go out from the finite size scaling region. We can tentatively
write $L/\xi \lesssim 3$, though this value is somewhat arbitrary. In this 
case this choice corresponds to eliminate the final four points in
Fig.~\ref{scaising.fig}. Condition~\eq{cond1} eliminates the five
points closer to $s=0$ in that figure. Putting the two conditions
together, the resulting lattices which should show a proper scaling
are given in the third column of Table~\ref{table:xi}. These are the
middle section, 6th-20th points in Fig.~\ref{scaising.fig}, which indeed 
seem to lie on a single curve.
 
As a further check, we can fit the critical exponent $\delta$
by means of Eq.~(\ref{rhoisingsca}) using as an input  
$\nu$ and $\bc$ ($\nu=1$, $\bc=0.44068...$).
For the unknown term depending on $f(s)$ we have several
possibilities. We can guess that it is 
roughly constant inside the range of the explored $s$, and do 
the fit to (fit (a))
\begin{equation}
\rho_{\textrm{rescal}}=-\frac{4\delta}{s}\ ,
\label{fita}
\end{equation}
or consider that when $s\to 0$, both $f(s)$ and its derivative go
to a constant, and write (fit (b))
\begin{equation}
\rho_{\textrm{rescal}}=-\frac{4\delta}{s} + C \ ,
\end{equation}
where $C$ is a constant term. We can finally use the three-parameter fit
(fit (c))
\begin{equation}
\rho_{\textrm{rescal}}=-\frac{4\delta\,(s+a)}{(s+a)^2 + b^2} \ .
\label{fitc}
\end{equation}
 
\begin{table}[tb]
\begin{center}
\begin{tabular}{ccccccc}\hline
 & \multicolumn{2}{c}{Fit (a)} & \multicolumn{2}{c}{Fit (b)} &
\multicolumn{2}{c}{Fit (c)} \\ \hline\hline
Points & $\delta$ & $\chi^2$/DF & $\delta$ & $\chi^2$/DF & $\delta$ & 
$\chi^2$/DF \\
\hline 
6-20 & 0.120(5) & 1.42 & 0.135(2) & 1.21 & 0.132(10) & 0.56 \\
7-20 & 0.120(6) & 1.53 & 0.136(2) & 1.22 & 0.128(9) & 0.36 \\
8-20 & 0.110(3) & 0.52 & 0.143(3) & 0.48 & 0.120(13) & 0.36 \\
\hline
\end{tabular}
\end{center}
\caption{Different fits to obtain the $\delta$ exponent from 
Eqs.~\eq{fita}--\eq{fitc} (see text).}
\label{table:fits}
\end{table}

The results of the fits\footnote{The fits have been performed by using the
MINUIT program.} are given in Table~\ref{table:fits}. Eliminating the 6th
or the 7th point (second and third rows in Table~\ref{table:fits}) corresponds
to taking Eq.~\eq{cond1} as a rigourous inequality, $L>\xi$, according to
the data of Table~\ref{table:xi}. Fit (c) gives the more stable values
with respect to the modification of the number of points in the fit, and
also those with less $\chi^2$ by degree of freedom (DF). 
A mean of the values given by the fit (c)
(third column in Table~\ref{table:fits}) gives $\delta = 0.127(8)$, 
in fair agreement with the expected value $\delta = 0.125$. 
This fit for points 7-20 is shown in Fig.~\ref{scaising.fig}.

A more complicated fit allowing also the determination
of $\nu$ and $\bc$ in principle is also possible, but an appropriate
guess of the form of $f$ is needed (see 
\cite{paffuti,mario,dario}).
This fit is useful when only approximated values of $\bc$ and $\nu$ are
known. In our case the solution of the model is known
and the simple quality of the scaling is itself a
good indication of the correctness of the approach. 
We note however that in order to obtain an accurate result for the critical
exponents, one should not approach very much the critical
point, where the correlation length diverges, in contrast with 
ordinary analyses. This can be taking into
account by monitoring the correlation length, as we showed in this
simple case.

\section{Conclusions}
\label{sez4}
The analysis of the Ising model suggests that 
the study of $\rho$ is suitable for investigating order-disorder phase 
transitions from a dual point of view. While
$\pardis$ is not adequate for numerical simulations, $\rho$ data are
completely reliable and allow an unambiguous reconstruction of the shape of
$\pardis$. In addition, from a finite size scaling
analysis we can get the critical
exponent associated to the disorder parameter, the critical
temperature and the critical exponent of the correlation length,
though a careful analysis for this observable is convenient
in order to get an accurate measure for the critical exponents.

This analysis is the first complete test of the procedure used
to demonstrate that dual superconductivity of the vacuum is
the mechanism of colour confinement. The result of this test
confirms the hypotheses that are at the basis of the study of
Ref.~\cite{digiaco}. 

\begin{ack}
This work has been partially supported by the EC TMR program
ERBFMRX-CT97-0122, and by the Italian MURST. B.L. is supported
by the PPARC grant PPA/G/0/1998/00567 {\em Theoretical studies
of elementary particles and their interactions}.
\end{ack}

\end{document}